\documentclass[
superscriptaddress,
nofootinbib,
onecolumn,
 amsmath,amssymb,
 aps,
pra,
notitlepage,
longbibliography
]{revtex4-1}
\usepackage[pdftex=true,linktoc=page]{hyperref} 
\usepackage{dcolumn}
\usepackage{bm}
\usepackage{epsfig}
\usepackage{graphicx}
\usepackage{latexsym}
\usepackage{amsfonts}
\usepackage{setspace}
\usepackage{graphicx}
\usepackage{amsmath}
\usepackage{verbatim}
\usepackage{color}
\usepackage{SIunits}
\usepackage{hyperref}
\usepackage{float}

\newcommand{\be}{\begin{equation}}
\newcommand{\ee}{\end{equation}}
\newcommand{\ba}{\begin{array}}
\newcommand{\ea}{\end{array}}
\newcommand{\bqa}{\begin{eqnarray}}
\newcommand{\eqa}{\end{eqnarray}}
\newcommand{\bd}{\bold}

\let\vec\boldsymbol

\begin{document}

\title{Mechanical bound states in the continuum for macroscopic optomechanics}

\author{Mengdi Zhao} 
\affiliation{Department of Physics, University of Illinois at Urbana-Champaign, Urbana, IL 61801 USA}
\affiliation{Micro and Nanotechnology Laboratory, University of Illinois at Urbana-Champaign, Urbana, IL 61801 USA}
\author{Kejie Fang} 
\email{kfang3@illinois.edu}
\affiliation{Micro and Nanotechnology Laboratory, University of Illinois at Urbana-Champaign, Urbana, IL 61801 USA}
\affiliation{Department of Electrical and Computer Engineering, University of Illinois at Urbana-Champaign, Urbana, IL 61801 USA}

\begin{abstract} 
Bound states in the continuum (BICs), an emerging type of long-lived resonances different from the cavity-based ones, have been explored in several classical systems, including photonic crystals and surface acoustic waves. Here, we reveal symmetry-protected mechanical BICs in the structure of slab-on-substrate optomechanical crystals. Using a group theory approach, we identified all the mechanical BICs at the $\Gamma$ point in optomechanical crystals with $C_{4v}$ and $C_{6v}$ symmetries as examples, and analyzed their coupling with the co-localized optical BICs and guided resonances due to both moving boundary and photo-elastic effects. We verified the theoretical analysis with numerical simulations of specific optomechanical crystals which support substantial optomechanical interactions between the mechanical BICs and optical resonances. Due to the unique features of high-$Q$, large-size mechanical BICs and substrate-enabled thermal dissipation, this architecture of slab-on-substrate optomechanical crystals might be useful for exploring macroscopic quantum mechanical physics and enabling new applications such as high-throughput sensing and free-space beam steering.

\end{abstract}

\maketitle

\section{\label{sec:level1}Introduction}

Light trapping is a key technique for enhancing light-matter interactions in physical architectures such as cavity-QED \cite{walther2006cavity} and cavity-optomechanics \cite{aspelmeyer2014cavity}. One method to achieve light confinement at wavelength scales is by dispersion engineering of light to create localized optical resonances which are spectrally separated from the radiation continuum, as represented by the photonic crystal cavities. Recently, a different mechanism for light trapping has gained resurgent interest, where long-lived optical resonances can reside in the radiation continuum while still being confined in certain spatial dimensions \cite{hsu2016bound}. In contrast to conventional optical cavities with hard ``mirrors'', such bound states in the continuum (BICs) can exist simply due to the symmetry incompatibility between the confined modes and the free-space modes, decoupling the former from the radiation continuum. While for other non-symmetry-protected BICs, it generally requires tuning of system parameters to accidentally cancel all the radiation amplitudes.

Although the concept of BICs can be traced back to the early work by Wigner and von Neumann in quantum mechanics \cite{von1929some}, they are now more commonly found in classical systems. For example, in photonic crystal slabs, both symmetry-protected and non-symmetry-protected optical BICs have been observed \cite{lee2012observation, hsu2013observation}, leading to applications including large-area lasing \cite{kodigala2017lasing}. Certain types of surface acoustic waves in anisotropic elastic media or layered structures are also identified as BICs \cite{lim1969character,maznev2018bound}. However, acoustic BICs in artificial periodic structures, such as mechanically-compliant photonic crystals, have not been explored. In mechanically-compliant photonic crystals, acoustic modes can strongly interact with the optical modes via radiation-pressure force, leading to the paradigm of optomechanical crystals \cite{eichenfield2009optomechanical} which are widely explored for inertia sensing \cite{krause2012high}, information processing \cite{fang2016optical}, and quantum science \cite{riedinger2018remote}. In contrast to most free-standing optomechanical structures, here we study slab-on-substrate optomechanical crystals, in which mechanical BICs can exist due to symmetry incompatibility with acoustic waves in the substrate, and analyze their interactions with the co-localized optical resonances using a group theory approach. We verify the theoretical analysis using numerical simulations of specific types of optomechanical crystals which exhibit substantial coupling between the mechanical BICs and optical resonances.

This new optomechanical architecture, with long-lived mechanical modes in the continuum, might help resolve some long-standing problems in cavity-optomechanics. Conventional on-chip optomechanical structures are suspended to prevent leakage of resonance phonons into the substrate. However, such suspended structures also result in slow heat dissipation which limits the parametric pump power and causes excess noise, especially when they are operated at low temperature and near quantum regime \cite{cohen2015phonon,meenehan2015pulsed}. The structure of slab-on-substrate optomechanical crystals, while trapping phonons of specific frequencies in mechanical BICs, facilitates the dissipation of thermal phonons as well as other non-resonant phonons via direct contact with the substrate. This mechanism is expected to reduce optically induced thermal noise and nonlinear acoustic noise while accommodating optical pumps for parametrically-enhanced coupling between the macroscopic mechanical BICs and optical resonances, for optically cooling the mechanical modes and addressing single phonons \cite{riedinger2018remote}. 

In principle, the mechanical BICs in optomechanical crystal slabs are not limited in size in the planar dimensions, in contrast to the resonances in micro-cavities which have a mode size comparable to the wavelength. Thus, such mechanical BICs with large masses, as well as their frequencies in the gigahertz range, represent a unique system for studying macroscopic quantum physics. The large-area mechanical BICs coupled with optical fields might also enable new applications, such as high-throughput sensing and free-space beam steering.

\section{\label{sec:level2}Symmetry-protected mechanical and optical BICs in slab-on-substrate optomechanical crystals}

In photonic crystal slabs with a spatial symmetry represented by a point group $G$, the optical modes with Bloch momentum $\bd k$ belong to the irreducible representation of the $\bd k$-group $G_{\bd k}$, i.e., a sub-group of $G$ whose elements keep vector $\bd k$ invariant \cite{inui2012group}. When the representation of a photonic crystal mode is incompatible with that of far-field radiation modes with the same parallel momentum, this mode decouples from the radiation continuum and becomes trapped in the photonic crystal slab \cite{ochiai2001dispersion}. In mechanically-compliant photonic crystals, i.e., optomechanical crystals, the  mechanical Bloch modes can also be classified by the irreducible representations of the point group, and symmetry-protected mechanical BICs exist when their representations are incompatible with the acoustic waves in the surrounding media. We are interested in the coupling between mechanical BICs and optical resonances, including BICs and guided resonances, in slab-on-substrate optomechanical crystals. In general, because of the phase matching condition, only mechanical modes with Bloch momentum $\bd k=0$ can couple with an optical mode. As such, we will only consider mechanical BICs at the $\Gamma$ point in this paper, whose $\bd k$-group is simply $G$. However, mechanical BICs at other high-symmetric points in the Brillouin zone can be similarly analyzed, which are useful for coupling two optical modes with different momenta.

\begin{figure}[!htb]
\begin{center}
\includegraphics[width=0.4\columnwidth]{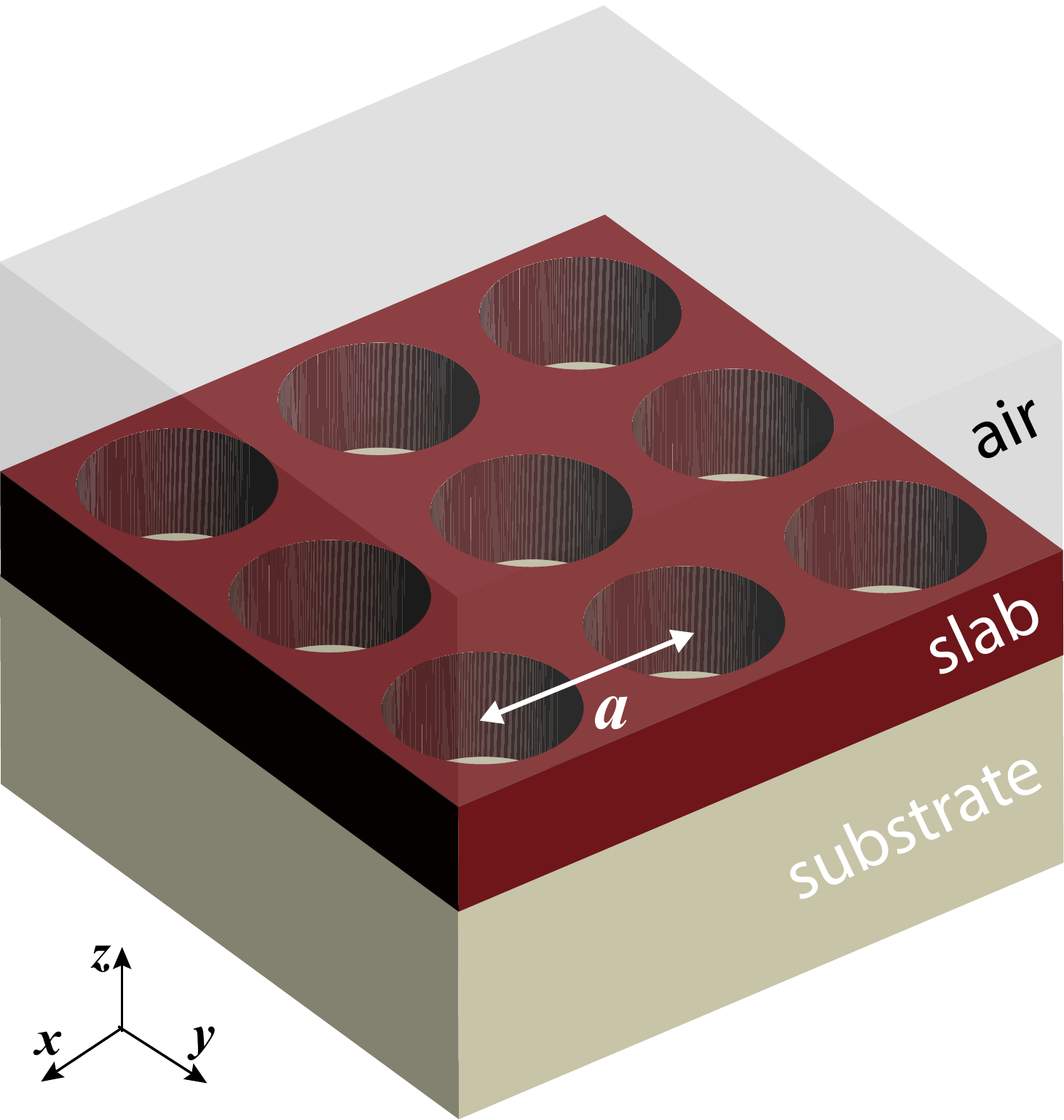}
\caption{Schematic of a slab-on-substrate optomechancial crystal with $C_{4v}$ symmetry. }
\label{fig:lattice}
\end{center}
\end{figure}

As examples, we will analyze two types of slab-on-substrate optomechanical crystals, with spatial symmetry described by $C_{4v}$ and $C_{6v}$ groups, which are commonly seen in crystals with square and triangular lattices, respectively. The structure with $C_{4v}$ symmetry is schematically shown in Fig. \ref{fig:lattice} with a square lattice (lattice constant $a$) of circular holes in the slab, which is sandwiched between the semi-infinite substrate and air. The specific structure of the unit cell is for the purpose of illustration, which can be of any geometry that respects the point group symmetry for the analysis in this paper to apply. In Table \ref{tbl:c4v} and \ref{tbl:c6v}, we list all the irreducible representations of the $C_{4v}$ and $C_{6v}$ groups and their characters to facilitate the discussion below.

\begin{table}[ht]
\parbox{.45\linewidth}{
\caption{Character table for the $C_{4v}$ point group}
\centering
\begin{tabular}{c |c c c c c} 
\hline\hline
$C_{4v}$ & $E$ & $2C_4$ & $C_2$ & $2\sigma_v$ & $2\sigma_d$ \\ 
\hline 
$A_1$ & 1 & 1 & 1 & 1 & 1 \\ 
$A_2$ & 1 & 1 & 1 & -1 & -1\\
$B_1$ & 1 & -1 & 1 & 1 & -1\\
$B_2$ & 1 & -1 & 1 & -1 & 1\\
$E$ & 2 & 0 & -2 & 0 & 0 \\ 
\hline
\end{tabular}
\label{tbl:c4v}
}
\hfill
\parbox{.45\linewidth}{
\caption{Character table for the $C_{6v}$ point group}
\centering
\begin{tabular}{c |c c c c c c} 
\hline\hline
$C_{6v}$ & $E$ & $2C_6$ & $2C_3$ & $C_2$ & $3\sigma_y$ & $3\sigma_x$ \\ 
\hline 
$A_1$ & 1 & 1 & 1 & 1 & 1 & 1 \\ 
$A_2$ & 1 & 1 & 1 & 1 & -1 & -1\\
$B_1$ & 1 & -1 & 1 & -1 & 1 & -1\\
$B_2$ & 1 & -1 & 1 & -1 & -1 & 1\\
$E_1$ & 2 & 1 & -1 & -2 & 0 & 0 \\
$E_2$ & 2 & -1 & -1 & 2 & 0 & 0 \\ 
\hline
\end{tabular}
\label{tbl:c6v}
}
\end{table}

\subsection{Mechanical BICs}

The evolution of elastic waves in solids is governed by the following wave equation of elastodynamics:
\begin{equation}
\rho\frac{\partial^2\vec{Q}}{\partial t^2}=\frac{E}{2(1+\nu)}\nabla ^2\vec{Q}+\frac{E}{2(1+\nu)(1-2\nu)}\nabla\nabla\cdot\vec{Q},
\label{eqn:elastic}
\end{equation}
where $\rho$, $E$ and $\nu$ are the density, Young's modulus and Poisson's ratio of the solid, respectively, and $\vec{Q}$ is the mechanical displacement. According to Bloch's theorem, when the elastic parameters are periodically distributed in two dimensions, such as in optomechanical crystal slabs, the mechanical eigenmodes of Eq. \ref{eqn:elastic} can be written as:
\begin{equation}\label{BlochQ}
\vec{Q}_{\vec{k}_\parallel}(\vec{r})e^{-i\omega t}=\vec{u}_{\vec{k}_{\parallel}}(\vec{r})e^{i(\vec{k}_{\parallel}\cdot \vec{\rho}-\omega t)},
\end{equation}
where $\omega$ is the eigenfrequency, $\vec{k}_{\parallel}$ is the Bloch wavevector in the plane of the slab, $\vec{\rho}$ is the in-plane component of the position vector $\bd r$, and $\vec{u}_{\vec{k}_{\parallel}}(\vec{r})$ is a periodic function in the two dimensions. We can further expand $\vec{u}_{\vec{k}_{\parallel}}(\vec{r})$ using discrete Fourier transform:
\begin{equation}
\vec{u}_{\vec{k}_{\parallel}}(\vec{r})=\sum_{j=0}^{\infty}\vec{f}_j(z)e^{i\vec{G}^j_\parallel\cdot\vec{\rho}},
\label{eqn:elasticfourier}
\end{equation}
where $\vec{G}_{\parallel}^j$ are the reciprocal lattice vectors. For example, in an optomechanical crystal with a square lattice, the first few reciprocal lattice vectors are $\vec{G}_{\parallel}^0=(0,0)\frac{2\pi}{a}$, $\vec{G}_{\parallel}^{j=1,2,3,4}=(\pm 1,0)\frac{2\pi}{a}$ and $(0,\pm 1)\frac{2\pi}{a}$, etc.. 

In the far field, the solution of $\vec{Q}_{\vec{k}_{\parallel}}(\vec{r})$ is merely a superposition of plane waves. To find it, we can decompose $\vec{Q}$ into the curl-free ($\vec{Q}_L$) and divergence-free ($\vec{Q}_T$) components and separate Eq. \ref{eqn:elastic} into two independent equations:
\begin{equation}
\frac{\partial^2\vec{Q}_{T(L)}}{\partial t^2}=c_{T(L)}^2\nabla^2\vec{Q}_{T(L)},
\label{eqn:elastic2}
\end{equation}
where $c_T=\sqrt{\frac{E}{\rho}\frac{1}{2(1+\nu)}}$ and $c_L=\sqrt{\frac{E}{\rho}\frac{1-\nu}{(1-2\nu)(1+\nu)}}$ are the speed of transverse and longitudinal elastic waves, respectively. Note the elastic parameters in these formula of speed of sound are those of the substrate. The mechanical mode in the far field is then given by 
\begin{equation}
\vec{Q}_{\vec{k}_\parallel,\infty}(\vec{r})=e^{i\vec{k}_\parallel\cdot\vec{\rho}}\sum_{j=0}^{\infty}(\vec{A}_{T,j}e^{ik^j_{T,z}z}+\vec{A}_{L,j}e^{ik^j_{L,z}z})e^{i\vec{G}^j_\parallel\cdot\vec{\rho}},
\label{eqn:phononicplanewave}
\end{equation}
where 
\begin{equation}
k_{T(L),z}^j=\sqrt{\frac{\omega^2}{c_{T(L)}^2}-|\vec{k}_{\parallel}+\vec{G}_{\parallel}^j|^2},
\end{equation}
and $\vec{A}_{T,j}$ and $\vec{A}_{L,j}$ are the amplitudes of transverse and longitudinal waves, respectively.

Because of the phase-matching condition for non-vanishing optomechanical coupling between a mechanical mode and an optical mode, we only consider mechanical BICs at the $\Gamma$ point, i.e., $\vec{k}_{\parallel}=0$. In most solids, Poisson's ratio is less than one, leading to $c_T<c_L$. For $\vec{k}_{\parallel}=0$, and $\omega<c_T\frac{2\pi}{a}$ (for $C_{4v}$ structures) or $\omega<c_T\frac{4\pi}{\sqrt{3}a}$ (for $C_{6v}$ structures), which we call the cut-off frequency, only the zeroth-order terms ($j=0$) in Eq. \ref{eqn:phononicplanewave} have real $z$-component of the wavevector, and they can be parameterized as
\begin{equation}
\vec{Q}_0=\vec{Q}_{T,0}+\vec{Q}_{L,0}=(u\vec{e}_x+v\vec{e}_y)e^{ik_{T,z}^0z}+w\vec{e}_ze^{ik_{L,z}^0z}.
\end{equation}
These are the only radiation channels that a mechanical slab mode at the $\Gamma$ point below the cut-off frequency can couple to.

It is easy to see that $\{\vec{e}_x, \vec{e}_y\}$ forms the basis of the $E$ representation of $C_{4v}$ point group and the $E_1$ representation of $C_{6v}$ point group, and $\vec{e}_z$ belongs to the $A_1$ representation of $C_{4v}$ and $C_{6v}$ groups. Therefore, for a mechanical eigenmode with frequency $\omega<c_{T}\frac{2\pi}{a}$ in an optomechanical crystal with $C_{4v}$ symmetry, it can be a BIC only if its representation belongs to $\{A_2, B_1, B_2\}$. And in a structure with $C_{6v}$ symmetry, a mechanical mode with frequency $\omega<c_T\frac{4\pi}{\sqrt{3}a}$ can be a BIC only if its representation belongs to $\{A_2, B_1, B_2, E_2\}$. 

Above the cut-off frequency, at least the first-order transverse terms in Eq. \ref{eqn:phononicplanewave}, i.e., 
\begin{equation}
\vec{Q}_{1,T}=\sum_{l} \vec{A}_{T,l}e^{ik^{l}_{T,z}z}e^{i\vec{G}^{l}_\parallel\cdot\vec{\rho}},
\end{equation}
where $\{ \vec{G}_{\parallel}^{l}, l=1,\ldots,4\}=\{ \pm\frac{2\pi}{a}\vec{e}_{x,y}\}$ for $C_{4v}$ structures and $\{ \vec{G}_{\parallel}^{l}, l=1,\ldots,6\}=\{ \pm\frac{4\pi}{3a}\vec{e}_{x}, \pm\frac{4\pi}{3a}(\pm\frac{1}{2}\vec{e}_{x}+\frac{\sqrt{3}}{2}\vec{e}_{y})\}$ for $C_{6v}$ structures, become viable radiation channels for the mechanical modes at the $\Gamma$ point. We find the basis $\{ {\vec e_x}e^{i\vec{G}^{l}_\parallel\cdot\vec{\rho}}, {\vec e_y}e^{i\vec{G}^{l}_\parallel\cdot\vec{\rho}}\}$ can be decomposed into $A_1+B_1+A_2+B_2+2E$ representations of $C_{4v}$ group and $A_1+B_1+A_2+B_2+E_1+E_2$ representations of $C_{6v}$ group (see Appendix A), respectively, which contain all the irreducible representations of the two groups. As a result, above the cut-off frequency, no mechanical BICs exist. We summarize these results in Table \ref{tbl:bic}.

\begin{table}[ht]
\caption{Mechanical and optical BICs at the $\Gamma$ point} 
\centering 
\renewcommand{\arraystretch}{1.5}
\begin{tabular}{c | c | c } 
\hline\hline 
 & Mechanical BICs & Optical BICs \\ 
 & (only if $\frac{\omega}{2\pi}<\frac{c_T}{a} (C_{4v})  \textrm{ or }  \frac{2}{\sqrt{3}}\frac{c_T}{a} (C_{6v})$) & (only if $\frac{\omega}{2\pi}<\frac{c}{na}(C_{4v})$ or $ \frac{2}{\sqrt{3}}\frac{c}{na}(C_{6v})$) \\ 
\hline 
$C_{4v}$ & $A_2$, $B_1$, and $B_2$ & $A_1$, $A_2$, $B_1$, and $B_2$  \\ 
\hline
$C_{6v}$ & $A_2$, $B_1$, $B_2$, and $E_2$ & $A_1$, $A_2$, $B_1$, $B_2$, and $E_2$  \\ 
\hline 
\end{tabular}
\label{tbl:bic} 
\end{table}

\subsection{Optical BICs}
The analysis of optical BICs largely follows from the analysis of mechanical BICs above. Again, in order to have any confined optical modes in the slab, we require the refractive index of the material of slab is larger than that of the substrate. In the far field, the electric fields satisfy the Maxwell's equation,
\begin{equation}
\frac{1}{c^2}\frac{\partial^2\vec{E}}{\partial t^2}=\frac{1}{n^2}\nabla^2\vec{E},
\end{equation}
where $n$ is the refractive index of the substrate and $c$ is the speed of light in vacuum. We can decompose the electric fields into plane waves:
\begin{equation}
\vec{E}_{\vec{k}_\parallel}(\vec{r})=e^{i\vec{k}_\parallel\cdot\vec{\rho}}\sum_{j=0}^{\infty}\vec{A}_je^{ik^j_zz+i\vec{G}^j_\parallel\cdot\vec{\rho}},
\label{eqn:fourier}
\end{equation}
where
\begin{equation}
k^j_z=\sqrt{\frac{\omega^2}{(c/n)^2}-|\vec{k_\parallel}+\vec{G}^j_\parallel|^2}.
\end{equation}
The difference from elastic waves is that electromagnetic waves can only be transverse.

At the $\Gamma$ point, when $\omega<\frac{c}{n}\frac{2\pi}{a}$ ($C_{4v}$ structures) or $\omega<\frac{c}{n}\frac{4\pi}{\sqrt{3}a}$ ($C_{6v}$ structures), the only radiation component in Eq. \ref{eqn:fourier} is $\vec{E}_0=\vec{A}_0e^{ik_z^0z}$. $\vec{E}_0$ has the same form as $\vec{Q}_{T,0}$, so it belongs to the $E$ representation of $C_{4v}$ or the $E_1$ representation of $C_{6v}$. Above the cut-off frequencies, similar to the mechanical case, the radiation terms contain all the irreducible representations of $C_{4v}$ or $C_{6v}$ group, and thus no optical BICs exist. The result of optical BICs is summarized in Table \ref{tbl:bic} as well. 

\section{Optomechanical couplings between mechanical BICs and optical modes}

The optomechanical interaction between a mechanical mode and an optical mode can be modeled by the Hamiltonian $H=g_0a^\dagger a(b+b^\dagger)$, where $a(b)$ and $a^\dagger(b^\dagger)$ are the annihilation and creation operators of the optical(mechanical) mode, and $g_0$ is the single-photon optomechanical coupling. The main contribution to the optomechanical coupling includes moving boundary effect $g_{\textrm{0,MB}}$ and photo-elastic effect $g_{\textrm{0,PE}}$, with the total optomechanical coupling given by \cite{chan2012optimized}:
\begin{equation}\label{g0}
g_0=g_{0,\textrm{MB}}+g_{0,\textrm{PE}}\equiv\sqrt{\frac{\hbar}{2m_{\textrm{eff}}\omega_m}}(g_{\textrm{OM,MB}}+g_{\textrm{OM,PE}}),
\end{equation}
where $\omega_m$ is the frequency of the mechanical mode and $m_{\textrm{eff}}=\int \rho |\vec{Q}|^2\mathrm{d} V$ is its effective mass with properly normalized $\vec Q$. 

For mechanical and optical Bloch modes in optomechanical crystals, the optomechanical coupling can be calculated using
\be\label{omc}
g_0=\frac{\bar g_{0}}{\sqrt{N}}\frac{1}{N}\sum_{n,m}e^{i\bd k\cdot(n\bd a_1+m\bd a_2)},
\ee
where $\bd k$ is the Bloch wavevector of the mechanical mode, $\bd a_{1,2}$ are the lattice vectors, $N$ is the number of unit cells in the optomechanical crystal slab, and $\bar g_0$ is the coupling in a single unit cell at $(n, m)=(0, 0)$. For large enough optomechanical crystals, from Eq. \ref{omc}, $g_0$ is zero unless $\bd k=0$, which is equivalent to the phase-matching condition. Thus, for the mechanical BIC at the $\Gamma$ point, its coupling with an optical mode is given by
\be
g_0=\frac{\bar g_{0}}{\sqrt{N}}.
\ee
Because of the scaling of $1/\sqrt{N}$, the single-photon optomechanical coupling between the delocalized mechanical BICs and optical modes is generally smaller than that in state-of-the-art optomechanical crystal cavities \cite{chan2012optimized}. However, because of the better unit-area photon capacity thanks to the substrate-enabled heat dissipation, parametrically enhanced optomechanical coupling, i.e.,
\be
G=g_0\sqrt{Nn_p}=\bar g_{0}\sqrt{n_p},
\ee 
where $n_p$ is the number of pump photons in a single unit cell, might be comparable or even better than the state-of-the-art, given sufficient pump powers can be supplied.

Below, we will systematically study the optomechanical coupling between a mechanical BIC and an optical mode, based on the symmetry of the modes indicated by their representations. This general analysis helps to identify modes with desired symmetries for non-vanishing optomechanical couplings which can be further optimized through numerical simulations.

\subsection{Moving boundary effect}
\label{sec:movingboundary_theory}
The optomechanical coupling due to moving boundary effect is given by\cite{chan2012optimized}
\begin{equation}
g_{\textrm{OM,MB}}=-\frac{\omega_o}{2}\frac{\int (\vec{Q}\cdot \hat{\vec n})(\Delta\epsilon |\vec{E}_{\parallel}|^2-\Delta\epsilon^{-1}|\vec{D}_{\perp}|^2) \mathrm{d} S}{\int \vec{E}^{*}\cdot \vec{D} \mathrm{d} V},
\label{Eq:mbeffect}
\end{equation}
\\
where $\omega_o$ is the frequency of the optical mode, $\hat{\vec n}$ is the unit vector normal to the boundary, $\parallel$ and $\perp$ denote the components parallel and perpendicular to the boundary, $\Delta\epsilon=\epsilon_{\textrm{int}}-\epsilon_{\textrm{ext}}$ and $\Delta\epsilon^{-1}=\epsilon^{-1}_{\textrm{int}}-\epsilon^{-1}_{\textrm{ext}}$ ($\epsilon_{\textrm{ext}}$ is the permittivity of the media which $\hat{\vec n}$ points to and $\epsilon_{\textrm{int}}$ is the permittivity of the media on the other side of the boundary.). We define $f\equiv\Delta\epsilon |\vec{E}_{\parallel}|^2-\Delta\epsilon^{-1}|\vec{D}_{\perp}|^2$ to simplify the discussion below.

We first consider the coupling between a one-dimensional (1-d) representation mechanical BIC and a 1-d representation optical mode. Apparently, for 1-d representation optical modes, $f$ is even under mirror operations $\sigma_\alpha$, i.e., $\{2\sigma_v,2\sigma_d\}$ in the $C_{4v}$ group and $\{3\sigma_x,3\sigma_y\}$ in the $C_{6v}$ group, since $\vec{E}_{\parallel}$ and $\vec{D}_{\perp}$ are either even or odd under these mirror operations. Here we used subscript $\alpha$ to denote the angle between the x-z plane and the mirror plane, e.g., $\sigma_{\pi/2}=\sigma_x$. The integral in the numerator of Eq. \ref{Eq:mbeffect} can be calculated in the two regions separated by the mirror plane,
\begin{equation}
\begin{aligned}
\int(\vec{Q}\cdot\hat{\vec n})f \, \mathrm{d} S&=\int_{\alpha}^{\alpha+\pi}(\vec{Q}\cdot\hat{\vec n})f \,\mathrm{d} S +\int_{\alpha}^{\alpha-\pi}(\vec{Q}\cdot\hat{\vec n})f \,\mathrm{d} S \\
&=\int_{\theta=0}^{\pi}\big(\vec{Q}(\alpha+\theta)+\sigma_{\alpha}\vec{Q}(\alpha-\theta)\big)\cdot\hat{\vec n}(\alpha+\theta)f(\alpha+\theta) \,\mathrm{d} S.
\end{aligned}
\label{Eq:mb_sigma}
\end{equation}
We see that, if the mechanical mode is odd under $\sigma_{\alpha}$, the integration is zero. According to Tables \ref{tbl:c4v} and \ref{tbl:c6v}, for both $C_{4v}$ and $C_{6v}$ groups, only $A_1$ is even under all mirror reflections. However, mechanical modes with $A_1$ representation cannot be a BIC as shown above. As a result, the interaction between a 1-d mechanical BIC and a 1-d optical mode due to moving boundary effect is always zero. 

Next we consider the case when at least one of the mechanical and optical modes is a 2-d representation. For a 2-d representation, the matrix form of the group elements are generally not diagonal, except for $C_2$. One can find $C_2=-I$ for $E\ (C_{4v})$ and $E_1\ (C_{6v})$ representations and $C_2=I$ for $E_2\ (C_{6v})$ representation. Therefore, all the modes, irrespective of the dimension of the representation, are either odd or even under $C_2$. Making use of the unique property of $C_2$, the integral in the numerator of Eq. \ref{Eq:mbeffect} can be calculated as
\begin{equation}
\begin{aligned}
\int(\vec{Q}\cdot\hat{\vec n})f \, \mathrm{d} S&=\int_{0}^{\pi}(\vec{Q}\cdot\hat{\vec n})f \,\mathrm{d} S+\int_{\pi}^{2\pi}(\vec{Q}\cdot\hat{\vec n})f \,\mathrm{d} S \\
&=\int_{\theta=0}^{\pi}(\vec{Q}(\theta)+C_2\vec{Q}(\theta+\pi))\cdot\hat{\vec n}(\theta)f(\theta) \,\mathrm{d} S.
\end{aligned}
\label{Eq:mb_c2}
\end{equation}
From Eq. \ref{Eq:mb_c2}, we see that the mechanical BIC needs to be even under $C_2$ for non-vanishing moving boundary effect in this case.

The optomechanical coupling due to moving boundary effect between a mechanical BIC and an optical mode is summarized in Table \ref{tbl:mbsquare} and Table \ref{tbl:mbtriangular} for structures with $C_{4v}$ and $C_{6v}$ symmetry, respectively. In these tables, when the coupling is indicated as $\neq 0$, it only means the coupling is not constrained to be zero by symmetry; but could be zero due to other reasons, such as linear combination of degenerate modes.

\begin{table}[ht]
\caption{Moving boundary effect $g_{\textrm{OM,MB}}$, $C_{4v}$} 
\centering 
\renewcommand{\arraystretch}{1.5}
\begin{tabular}{c|c|c} 
\hline\hline 
  &  1-d optical BIC  & 2-d optical mode  \\
  & ($A_1$,$A_2$,$B_1$,$B_2$) & ($E$) \\
\hline 
1-d mechanical BIC & 0 & $\neq 0$ \\ 
($A_2$,$B_1$,$B_2$) & & \\
\hline 
\end{tabular}
\label{tbl:mbsquare} 
\end{table}

\begin{table}[ht]
\caption{Moving boundary effect $g_{\textrm{OM,MB}}$, $C_{6v}$} 
\centering 
\renewcommand{\arraystretch}{1.5}
\begin{tabular}{c | c | c } 
\hline\hline 
 &  1-d optical BIC  & 2-d optical BIC ($E_2$) \\
  & ($A_1$,$A_2$,$B_1$,$B_2$) & and other mode ($E_1$) \\
\hline 
1-d mechanical BIC & 0 & $\neq 0$ only for $A_2$ \\ 
($A_2$,$B_1$,$B_2$) & & \\
\hline
2-d mechanical BIC& $\neq 0$ & $\neq 0$  \\ 
 ($E_2$) & & \\
\hline 
\end{tabular}
\label{tbl:mbtriangular} 
\end{table}

\subsection{Photo-elastic effect}
\label{sec:photoelastic_theory}
The optomechanical coupling due to photo-elastic effect is given by\cite{chan2012optimized}
\begin{equation}
g_{\textrm{OM,PE}}=\frac{\omega_o}{2}\frac{\int \epsilon_0n^4 E_i^*E_j p_{ijkl}S_{kl} \,\mathrm{d} V}{\int \vec{E}^{*}\cdot \vec{D} \,\mathrm{d} V},
\label{Eq:pecoupling}
\end{equation}
where $\epsilon_0$ is the vacuum permittivity, $n$ is the refractive index, $p$ is the rank-four photo-elastic tensor and $S$ is the strain tensor. 

Unlike the moving boundary effect, the photo-elastic effect depends on the crystalline type of materials through the photo-elastic tensor. We will consider two types of crystalline: the hexagonal crystal, e.g., aluminum nitride, and the cubic crystal, e.g., silicon. Another subtlety of photo-elastic effect is the orientation of the material's crystal lattice relative to the artificial optomechanical crystal. To take this into account, we use a general photo-elastic tensor with an in-plane rotation with respect to the optomechanical crystal slab,
\begin{equation}
p'_{ijkl}(\theta)=R_{iq}(\theta)R_{jr}(\theta)R_{ks}(\theta)R_{lt}(\theta)p_{qrst},
\end{equation}
where
\begin{equation}
R(\theta)=\begin{pmatrix}
\textrm{cos}(\theta) & -\textrm{sin}(\theta) & 0\\
\textrm{sin}(\theta) & \textrm{cos}(\theta) & 0\\
0  &  0& 1\\
\end{pmatrix},
\end{equation}
and $\theta$ is the rotation angle. We find that the photo-elastic tensor of hexagonal crystals is independent of $\theta$, while that of cubic crystals is periodic in $\theta$ with a periodicity of $\pi/2$. The analysis of the photo-elastic effect is much more involved comparing to the moving boundary effect, due to a large number of terms present in Eq. \ref{Eq:pecoupling}. We leave the detailed analysis in Appendix B, and summarize the result in Tables \ref{tbl:pe_c4} and \ref{tbl:pe_c6}. In these Tables, when the coupling is indicated as $0\rightarrow \neq 0$, it means that the coupling depends on the rotation angle and it is 0 when $\theta$ is a multiple of $\pi/2$.

\begin{table}[ht]
\caption{Photo-elastic effect $g_{\textrm{OM,PE}}$, $C_{4v}$} 
\centering 
\renewcommand{\arraystretch}{1.5}
\begin{tabular}{c |c | c  } 
\hline\hline 
  &  1-d optical BIC  & 2-d optical mode  \\
 & ($A_1$,$A_2$,$B_1$,$B_2$) & ($E$)\\ 
\hline 
1-d mechanical BIC ($A_2$)& 0 (h); 0 $\rightarrow \neq 0$ (c)  & $ \neq 0$ \\ 
\hline
1-d mechanical BIC ($B_1$,$B_2$) & 0 & $ \neq 0$ \\
\hline 
\end{tabular}
\\
[1ex]
h: hexagonal crystal, c: cubic crystal, no indication: both.
\label{tbl:pe_c4} 
\end{table}

\begin{table}[ht]
\caption{Photo-elastic effect $g_{\textrm{OM,PE}}$, $C_{6v}$} 
\centering 
\renewcommand{\arraystretch}{1.5}
\begin{tabular}{c |c | c  } 
\hline\hline 
  & 1-d optical BIC  & 2-d optical BIC ($E_2$) \\
& ($A_1$,$A_2$,$B_1$,$B_2$) & and other mode ($E_1$)
 \\ 
\hline 
1-d mechanical BIC ($A_2$)& 0 (h); 0 $\rightarrow \neq 0$ (c)  & $\neq 0$\\ 
\hline
1-d mechanical BIC ($B_1$,$B_2$) & 0 & 0\\
\hline
2-d mechanical BIC ($E_2$) & $\neq 0$ & $\neq 0$\\
\hline 
\end{tabular}
\\
[1ex]
h: hexagonal crystal, c: cubic crystal, no indication: both.
\label{tbl:pe_c6} 
\end{table}

\section{\label{sec:simulation} Cross-structure optomechanical crystals: an example}

\begin{figure}[!htb]
\begin{center}
\includegraphics[width=0.8\columnwidth]{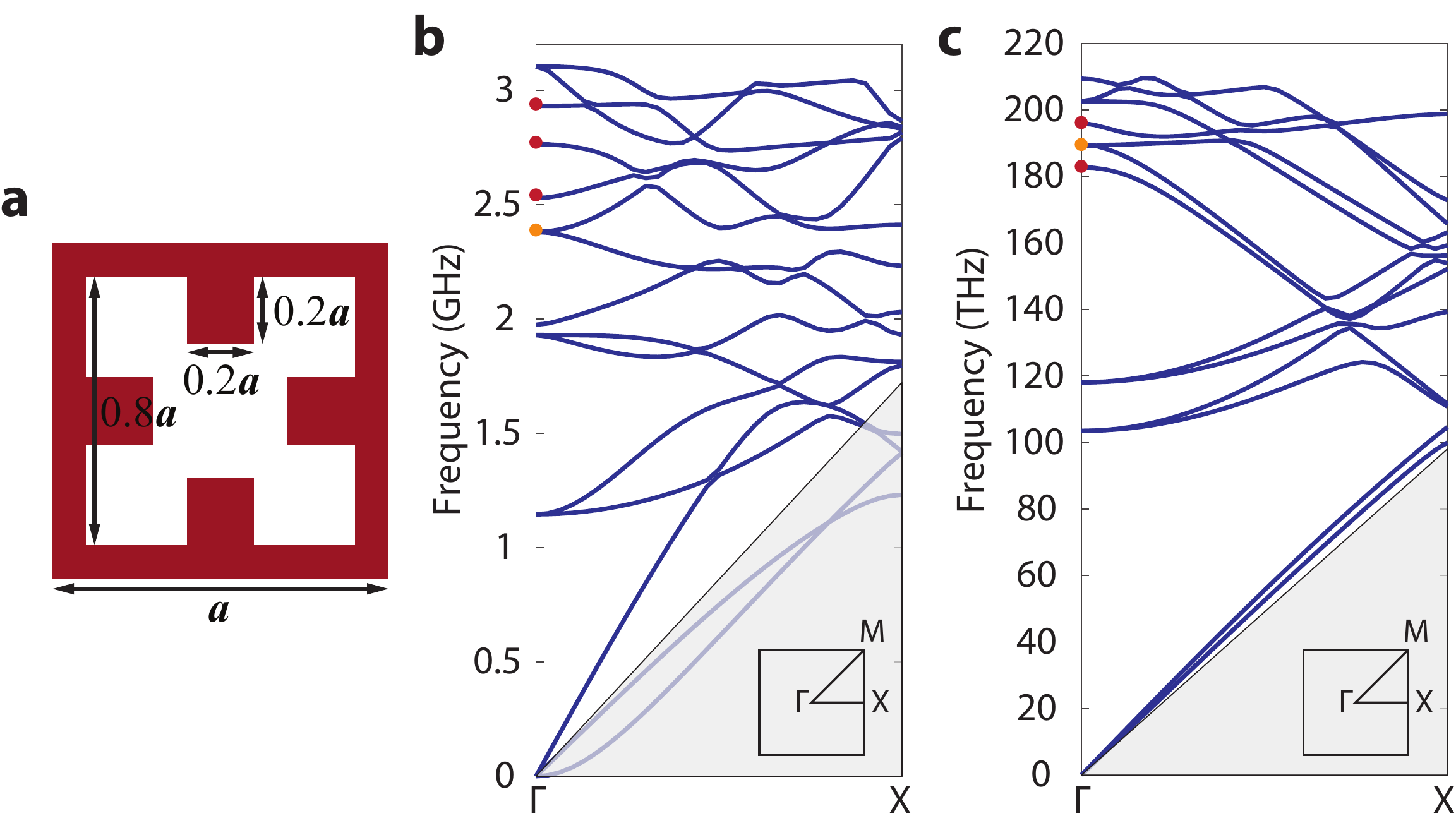}
\caption{\textbf{a}, Top view of the unit cell of an optomechanical crystal with $C_{4v}$ symmetry. $a=1\ \mu$m. \textbf{b}, The phononic bandstructure. The red dots indicate three mechanical BICs at the $\Gamma$ point and the yellow dot indicates a pair of degenerate mechanical guided resonances. \textbf{c}, The photonic bandstructure. The red dots indicate two optical BICs and the yellow dot indicates a pair of degenerate optical guided resonances. }
\label{fig:unit&bands}
\end{center}
\end{figure}

We designed and numerically simulated some slab-on-substrate optomechanical crystals with $C_{4v}$ and $C_{6v}$ symmetry to verify the theory above. For the structure with $C_{4v}$ symmetry, the unit cell of the optomechanical crystal is shown in Fig. \ref{fig:unit&bands}a. According to our simulation, such cross structures support more mechanical BICs than a unit cell with a regular square or circular hole. In our simulation, the slab is made of 600 nm thick aluminum nitride (AlN) and the substrate is silicon dioxide (SiO$_2$). For AlN, the Young's modulus, Poisson's ratio, density and refractive index are $325$ GPa, 0.25, 3120 kg/$m^3$ and 2.14, respectively; for SiO$_2$, these material parameters are $74.8$ GPa, 0.19, 2650 kg/$m^3$ and 1.53, respectively. Although the speed of sound in bulk AlN is greater than that of SiO$_2$, by making void optomechanical crystals in the slab, the effective speed of sound in the slab becomes smaller, which makes mechanical BICs below the cutoff frequency possible.

\begin{figure}[!htb]
\begin{center}
\includegraphics[width=0.8\columnwidth]{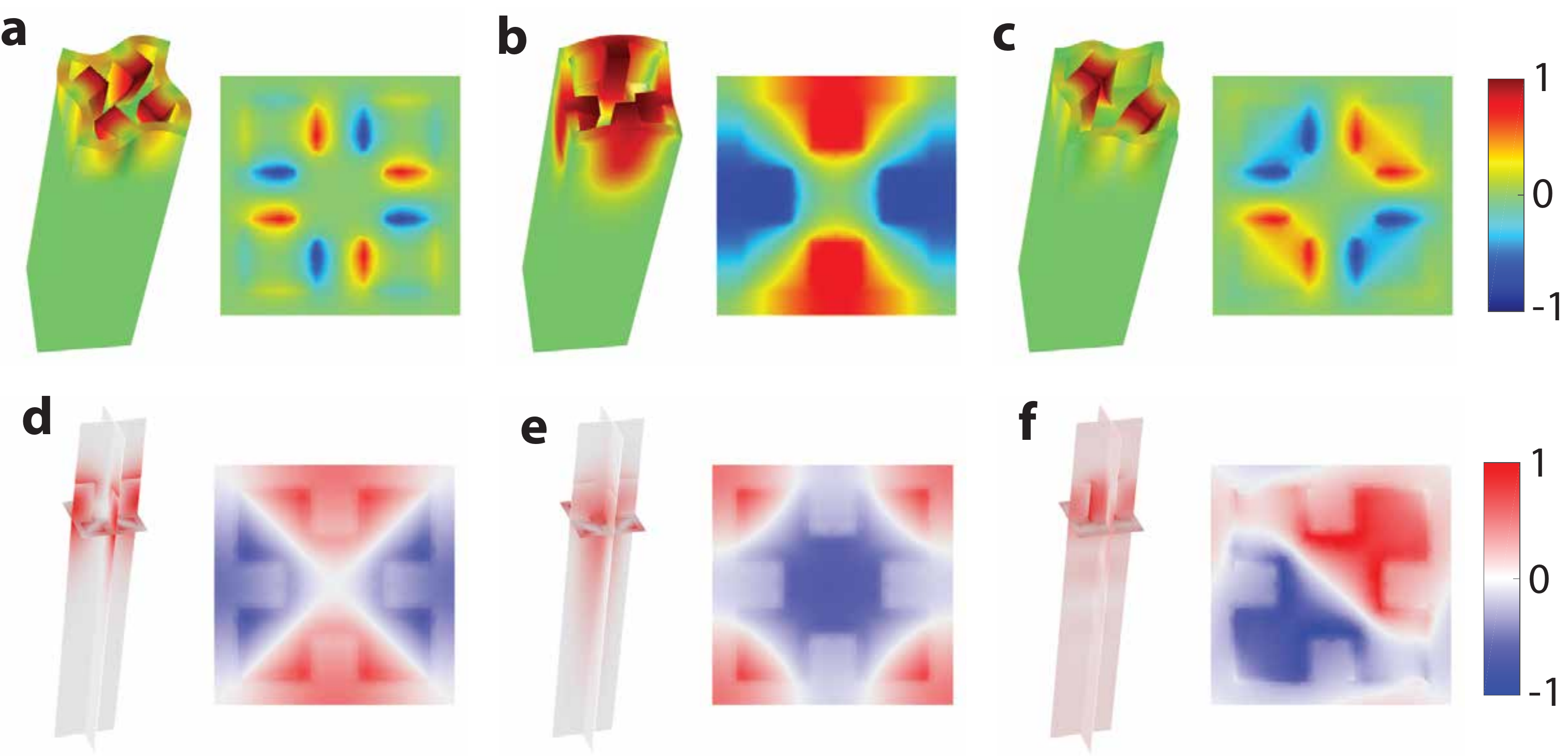}
\caption{Mechanical and optical modes at the $\Gamma$ point. \textbf{a}-\textbf{c}, Total displacement ($|\vec Q|$) (left) and $z$-component of the displacement ($Q_z$) at the interface of the slab and substrate (right) of the three mechanical BICs with frequency 2.53 GHz (\textbf{a}), 2.77 GHz (\textbf{b}) and 2.93 GHz (\textbf{c}). \textbf{d}-\textbf{f}, $|\vec{E}|^2$ (left) and $E_z$ (right) of the two optical BICs with frequency 182 THz (\textbf{d}) and 194 THz (\textbf{e}), and one of the degenerate guided resonance with frequency 190 THz (\textbf{f}).}
\label{fig:modes_square}
\end{center}
\end{figure}

\begin{figure}[!htb]
\begin{center}
\includegraphics[width=0.5\columnwidth]{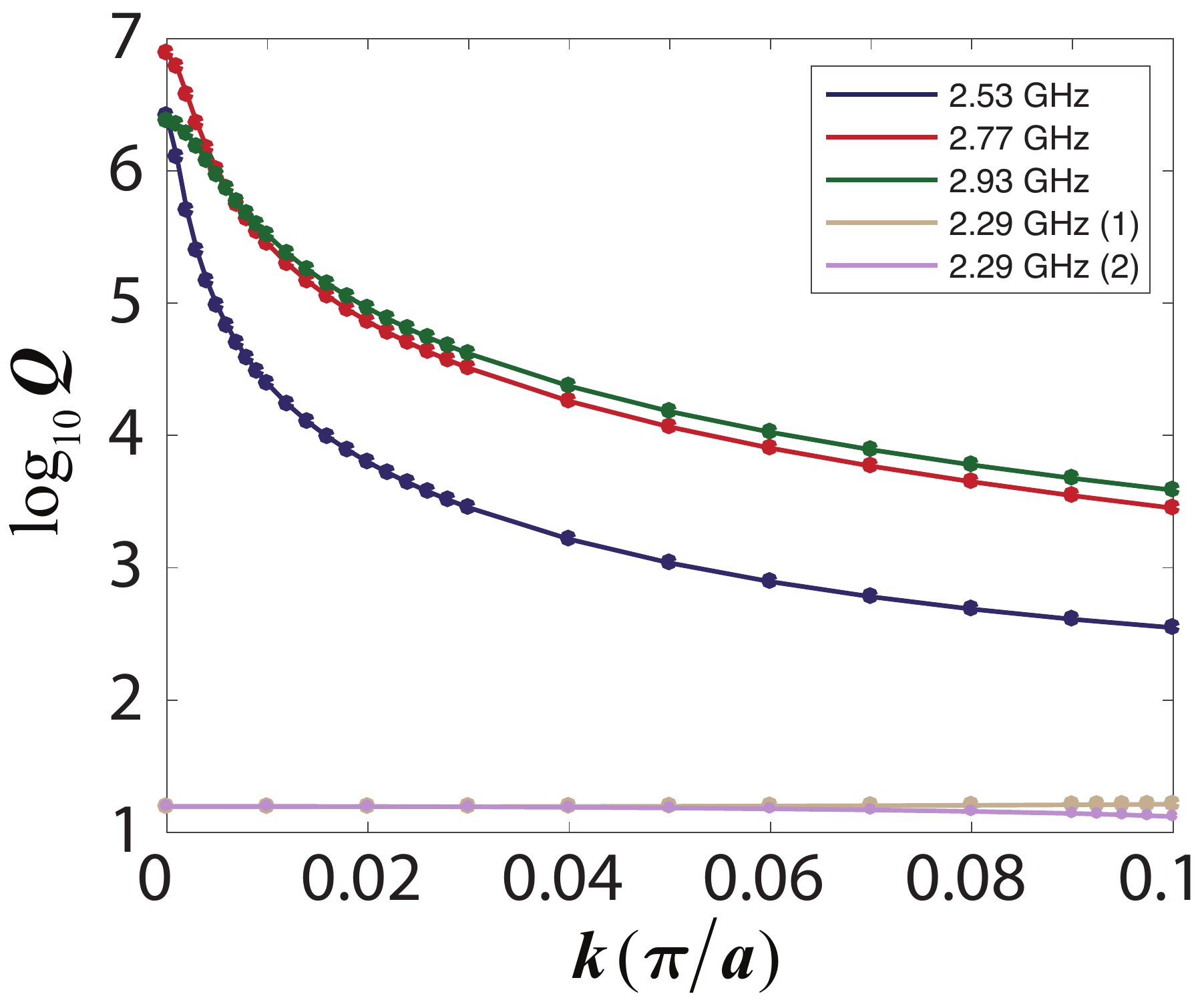}
\caption{Quality factor ($Q$) of the five mechanical modes (red and yellow dots in Fig. \ref{fig:unit&bands}b) as the Bloch wavevector is scanned near the $\Gamma$ point. The BICs have significantly higher quality factor than the guided resonances.}
\label{fig:phononicq}
\end{center}
\end{figure}

The phononic and photonic bandstructures for the unit cell with a lattice constant $a=1$ $\mu$m is shown in Fig.\ref{fig:unit&bands} b and c, calculated using finite element method (COMSOL) and plane wave expansion method (MIT Photonic Bands), respectively. We find that the modes at the $\Gamma$ point are consistent with the result of Table \ref{tbl:bic}. For the mechanical modes, below the cut-off frequency $c_{T,\textrm{SiO}_2}/a=$ 3.44 GHz, there are three mechanical BICs, indicated by the red dots in Fig. \ref{fig:unit&bands}b, with frequency and representation of (2.53 GHz, $A_2$), (2.77 GHz, $B_1$), and (2.93 GHz, $B_2$), respectively. For the optical modes, below the cut-off frequency $c/n_{\textrm{SiO}_2}a=196$ THz, there are two optical BICs, indicated by the red dots in Fig. \ref{fig:unit&bands}c, with frequency and representation of (182 THz, $B_1$) and (194 THz, $A_1$), respectively. The mode profiles of the three mechanical BICs and two optical BICs as well as an optical guided resonance (i.e., leaky modes which nevertheless have significant energy confined in the slab \cite{fan2002analysis}; the yellow dot in Fig. \ref{fig:unit&bands}c with frequency 190 THz and representation $E$) are shown in Fig. \ref{fig:modes_square}. Due to the finite thickness (4 $\mu$m) of the substrate used in our simulations, the mechanical BICs have limited quality factors instead of being infinity in theory. However, as shown in Fig. \ref{fig:phononicq}, comparing to the mechanical guided resonances (e.g. the yellow dot in Fig. \ref{fig:unit&bands}b), the quality factors of mechanical BICs are still significantly higher ($\sim 10^6-10^7$). 

We also calculated the optomechanical coupling between the mechanical BICs and optical modes. The photo-elastic tensor of hexagonal AlN is taken from Ref. \cite{davydov2002evaluation}. We found that, in this $C_{4v}$ structure, the optomechanical coupling between the mechanical BICs and optical BICs are indeed zero, and nonzero couplings occur between the mechanical BICs and the optical guided resonance with $E$ representation (Fig. \ref{fig:modes_square}f) as listed in Table \ref{tbl:square}, among which the largest total optomechanical coupling in a single unit cell is $\bar g_0/2\pi=690$ kHz for the mechanical BIC with frequency 2.93 GHz (Fig. \ref{fig:modes_square}c). 

\begin{table}[H]
\caption{Optomechanical couplings in a unit cell of the $C_{4v}$ cross-structure optomechanical crystal} 
\centering 
\renewcommand{\arraystretch}{1.5}
\begin{tabular}{c|c|c|c}   
\hline\hline 
 ($\bar g_{0,\textrm{MB}}/2\pi,\ \bar g_{0,\textrm{PE}}/2\pi$) (Hz) & 2.52 GHz ($A_2$)   & 2.77 GHz ($B_1$) & 2.93 GHz ($B_2$)\\
\hline 
 190 THz ($E$) & $(1.0\times 10^{3},-7.5\times 10^{2})$ &  $(7.7\times 10^{4},2.9\times 10^{2})$  & $(6.9\times 10^{5},4.0\times 10^{3})$  \\
\hline 
\end{tabular}
\label{tbl:square} 
\end{table}

We also simulated cross-structure optomechanical crystals with $C_{6v}$ symmetry in the same material system, and find that all the modes and optomechanical couplings are consistent with the theory. The result is provided in Appendix C. We note the optomechanical couplings obtained here are rather preliminary and can be further optimized with new structure design and parameter sweeping.

\section{SUMMARY}

In summary, we have proposed a new paradigm for chip-scale optomechanics using long-lived mechanical bound states in the continuum in slab-on-substrate optomechanical crystals. Using a group theory approach, we systematically identified mechanical BICs at the $\Gamma$ point in optomechanical crystals with $C_{4v}$ and $C_{6v}$ symmetry, and non-vanishing optomechanical couplings between the mechanical BICs and optical modes. We verified the theory with an example of slab-on-substrate optomechanical crystals that support a variety of mechanical BICs with large optomechanical couplings. The mechanical BICs in slab-on-substrate structures might be useful for exploring macroscopic quantum optomechanics and applications such as high-throughput sensing.


  
%

%

\appendix

\section{Representations of the first-order transverse radiation components}
\label{App:A}

In a square lattice, $\vec{Q}_{1,T}$ is given by
\begin{equation}
\vec{Q}_{1,T}=\sum_{l=0}^{3} \vec{A}_{T,1l}e^{ik^{1l}_{T,z}z}e^{i\vec{G}^{1l}_\parallel\cdot\vec{\rho}},
\end{equation}
where $\vec{G}_{\parallel}^{10}=\frac{2\pi}{a}\vec{e}_x$, $\vec{G}_{\parallel}^{11}=\frac{2\pi}{a}\vec{e}_y$, $\vec{G}_{\parallel}^{12}=-\frac{2\pi}{a}\vec{e}_x$, $\vec{G}_{\parallel}^{13}=-\frac{2\pi}{a}\vec{e}_y$. Since $\vec{Q}_{1,T}$ is divergence free, it can be expanded in the following basis: $\{\vec{q}_l=(C_4\vec{G}_{\parallel}^{1l})e^{i\vec{G}_{\parallel}^{1l}\cdot\vec{\rho}}\}_{l=0,...,3}\cup\{\vec{q}'_l=(\vec{G}_{\parallel}^{1l}/|\vec{G}_{\parallel}^{1l}|^2-\vec{e}_z/k_z)e^{i\vec{G}_{\parallel}^{1l}\cdot\vec{\rho}}\}_{l=0,...,3}$, where both $\vec q_l$ and $\vec q'_l$ are orthogonal to the wavevector $(\vec{G}^{1l}_\parallel, k^{1l}_{T,z}\vec e_z)$. We have omitted $e^{ik_zz}$ because it is invariant under any operation in the $C_{4v}$ point group.
We are going to prove the linear combination of $\{\vec{q}_l\}_{l=0,...,3}$ belongs to a reducible representation, which can be decomposed into $A_2+B_2+E$. Applying the operations in the $C_{4v}$ point group to $\{\vec{q}_l\}_{l=0,...,3}$, results in
\begin{equation}
    E\vec{q}_l=\vec{q}_l,\quad
    C_4\vec{q}_l=\vec{q}_{l+1},\quad
    C_2\vec{q}_l=\vec{q}_{l+2},\quad
    \sigma_x\vec{q}_l=-\vec{q}_{2-l},\quad
    \sigma_d'\vec{q}_l=\vec{q}_{1-l},
\end{equation}
with the subscripts in the sense of mod 4. Thus we get the characters of this reducible representation are:
\begin{equation}
    \chi(E)=4,\quad
    \chi(2\sigma_v)=-2,\quad
    \chi(2C_4)=\chi(C_2)=\chi(2\sigma_d)=0.
\end{equation}
Based on the character orthogonality theorem, the number of times that an irreducible representation appears in a reducible representation can be determined by
\begin{equation}
m_P=(\chi|\chi^{(P)})=\frac{1}{n}\sum_g\chi^{(P)}(g)^*\chi(g),
\end{equation}
where $P$ refers an irreducible representation, $\chi(g)$ and $\chi^{(P)}(g)$ are the characters of the operation g for the reducible representation and the irreducible representation $P$, respectively. The sum is over all the operations in the point group, whose number of elements is $n$. The numbers of irreducible representations are then:
\begin{equation}
    m_{A_1}=0,\quad
    m_{A_2}=1,\quad
    m_{B_1}=0,\quad
    m_{B_2}=1,\quad
    m_{E}=1.
\end{equation}
Thus any linear combination of $\{\vec{q}_l\}_{l=0,...,3}$ is a reducible representation that can be decomposed into the combination of an $A_2$ mode, a $B_2$ mode and a pair of $E$ modes. Likewise, 
\begin{equation}
    E\vec{q}'_l=\vec{q}'_l,\quad
    C_4\vec{q}'_l=\vec{q}'_{l+1},\quad
    C_2\vec{q}'_l=\vec{q}'_{l+2},\quad
    \sigma_x\vec{q}'_l=\vec{q}'_{2-l},\quad
    \sigma_d'\vec{q}'_l=\vec{q}'_{1-l},
\end{equation}
\begin{equation}
    \chi(E)=4,\quad
    \chi(2\sigma_v)=2,\quad
    \chi(2C_4)=\chi(C_2)=\chi(2\sigma_d)=0,
\end{equation}
\begin{equation}
    m_{A_1}=1,\quad
    m_{A_2}=0,\quad
    m_{B_1}=1,\quad
    m_{B_2}=0,\quad
    m_{E}=1.
\end{equation}
That is, the reducible representation of the linear combination of $\{\vec{q}'_l\}_{l=0,...,3}$ can be decomposed into $A_1+B_1+E$. Thus, $\vec{Q}_{1,T}$ can be decomposed into $A_1+B_1+A_2+B_2+2E$ in the square lattice. 

One can also verify that in the triangular lattice, the first order transverse radiation component
can be decomposed into $A_1+B_1+A_2+B_2+E_1+E_2$.

\section{Photo-elastic effect}
\label{App:B}

The photo-elastic matrices $P_h$ for hexagonal system ($D_{3h}$,$C_{6v}$,$D_6$,$D_{6h}$) and $P_c$ for cubic system ($T_d$,$O$,$O_h$) are as follows, 
\begin{equation}
P_h=\begin{pmatrix}
p_{11} &  p_{12} & p_{13} & 0 & 0 & 0\\
p_{12} & p_{11} & p_{13} & 0 & 0 & 0\\
p_{31} & p_{31} & p_{33} & 0 & 0 & 0\\
0 & 0 & 0 & p_{44} & 0 & 0\\
0 & 0 & 0 & 0 & p_{44} & 0\\
0 & 0 & 0 & 0 & 0 & \frac{1}{2}(p_{11}-p_{12})\\
\end{pmatrix},
\quad
P_c=\begin{pmatrix}
p_{11} &  p_{12} & p_{12} & 0 & 0 & 0\\
p_{12} & p_{11} & p_{12} & 0 & 0 & 0\\
p_{12} & p_{12} & p_{11} & 0 & 0 & 0\\
0 & 0 & 0 & p_{44} & 0 & 0\\
0 & 0 & 0 & 0 & p_{44} & 0\\
0 & 0 & 0 & 0 & 0 & p_{44}\
\end{pmatrix}.
\end{equation}
They can be generally described as:
\begin{equation}
P=\begin{pmatrix}
p_{11} &  p_{12} & p_{13} & 0 & 0 & 0\\
p_{12} & p_{11} & p_{13} & 0 & 0 & 0\\
p_{31} & p_{31} & p_{33} & 0 & 0 & 0\\
0 & 0 & 0 & p_{44} & 0 & 0\\
0 & 0 & 0 & 0 & p_{44} & 0\\
0 & 0 & 0 & 0 & 0 & p_{66}\\
\end{pmatrix}.
\end{equation}
Under the in-plane rotation:
\begin{equation}
R(\theta)=\begin{pmatrix}
\textrm{cos}(\theta) & -\textrm{sin}(\theta) & 0\\
\textrm{sin}(\theta) & \textrm{cos}(\theta) & 0\\
0  &  0& 1\\
\end{pmatrix},
\end{equation}
the photo-elastic tensor $P$ transforms according to 
\begin{equation}
p'_{ijkl}(\theta)=R_{iq}(\theta)R_{jr}(\theta)R_{ks}(\theta)R_{lt}(\theta)p_{qrst},
\end{equation}
which gives the non-zero elements in the rotated photo-elastic matrix $P'$:
\begin{equation}
\begin{array}{c}
p_{11}'=p_{22}'=\frac{1}{4}[p_{11}(\textrm{cos}(4\theta)+3)+(p_{12}+2p_{66})(1-\textrm{cos}(4\theta))]
\\
p_{33}'=p_{33}
\\
p_{12}'=p_{21}'=\frac{1}{4}[p_{12}(\textrm{cos}(4\theta)+3)+(p_{11}-2p_{66})(1-\textrm{cos}(4\theta))]
\\
p_{13}'=p_{23}'=p_{13}
\\
p_{31}'=p_{32}'=p_{31}
\\
p_{44}'=p_{55}'=p_{44}
\\
p_{66}'=\frac{1}{4}[2p_{66}(\textrm{cos}(4\theta)+1)+(p_{11}-p_{12})(1-\textrm{cos}(4\theta))]
\\
p_{16}'=p_{61}'=\frac{1}{4}\textrm{sin}(4\theta)(p_{11}-p_{12}-2p_{66})
\\
p_{26}'=p_{62}'=-p_{16}'.
\end{array}
\label{Eq:rotatedtensor}
\end{equation}
For the hexagonal crystal, $p_{66}=\frac{1}{2}(p_{11}-p_{12})$, which leads to $P'_h=P_h$, so $P_h$ is independent of $\theta$. While $P_c$ is periodic in $\theta$, with the periodicity being $\pi/2$.

Based on all the non-zero elements $p'_{ij}$ in the rotated matrix, the integration in the numerator of Eq. (18) in the main text can be written as
\begin{equation}
\begin{aligned}
\int -\epsilon_0n^4 E_i^*E_j p_{ijkl}S_{kl} \,\mathrm{d} V=\int& -\epsilon_0n^4 \big(p_{11}'f_{11}+p_{12}'f_{12}+p_{13}'f_{13}+p_{31}'f_{31}\\
&+p_{33}'f_{33}+p_{44}'f_{44}+p_{66}'f_{66}+p_{16}'f_{16}\big) \,\mathrm{d} V,
\end{aligned}
\end{equation}
where
\begin{equation}
\begin{array}{c}
f_{11}=|E_x|^2S_{xx}+|E_y|^2S_{yy}
\\
f_{12}=|E_x|^2S_{yy}+|E_y|^2S_{xx}
\\
f_{13}=|E_x|^2S_{zz}+|E_y|^2S_{zz}
\\
f_{31}=|E_z|^2S_{xx}+|E_z|^2S_{yy}
\\
f_{33}=|E_z|^2S_{zz}
\\
f_{44}=2Re\{E_y^*E_z\}\cdot 2S_{yz}+2Re\{E_x^*E_z\}\cdot 2S_{xz}
\\
f_{66}=2Re\{E_x^*E_y\}\cdot 2S_{xy}
\\
f_{16}=(|E_x|^2-|E_y|^2)\cdot 2S_{xy}+2Re\{E_x^*E_y\}\cdot (S_{xx}-S_{yy}).
\end{array}
\label{Eq:intterm}
\end{equation}

In Eq. (\ref{Eq:intterm}), if a term is odd under any of the operators, the integration of this term over the whole photo-elastic material domain will be zero, because the domain itself has $C_{4v}$ or $C_{6v}$ symmetry. TABLE \ref{tbl:underxy}, TABLE \ref{tbl:underdd} and TABLE \ref{tbl:underc2} list how $E_i^*E_j$ and $S_{ij}$ transform under each operation when the character of the operation is given. In these tables,``e" stands for even,``o" stands for odd and $\chi(R)$ denotes for the character of operation $R$.

\begin{table}[ht]
\renewcommand{\arraystretch}{1.5}
\caption{Transformations under $\sigma_x$ and $\sigma_y$}  
\centering 
\begin{tabular}{c | c  | c c c c c c} 
\hline\hline 
 & & $|E_x|^2$ & $|E_y|^2$ & $|E_z|^2$ & $Re\{E_x^*E_y\} $ & $Re\{E_x^*E_z\} $ & $Re\{E_y^*E_z\} $ \\
 \hline
$\chi(\sigma_x)=\pm 1$ & under $\sigma_x$ & e & e & e & o & o & e \\ 
 \hline
$\chi(\sigma_y)=\pm 1$ & under $\sigma_y$ & e & e & e & o & e & o \\ 
\hline\hline
 & & $S_{xx}$ & $S_{yy}$ & $S_{zz}$ & $S_{xy}$ & $S_{xz}$ & $S_{yz}$\\
\hline
$\chi(\sigma_x)=1$  & under $\sigma_x$ & e & e & e & o & o & e \\ 
\hline
$\chi(\sigma_x)=-1$  & under $\sigma_x$ & o & o & o & e & e & o \\ 
\hline
$\chi(\sigma_y)=1$ & under $\sigma_y$ & e & e & e & o & e & o \\ 
  \hline
$\chi(\sigma_y)=-1$ & under $\sigma_y$ & o & o & o & e & o & e  \\ 
\hline 
\end{tabular}
\label{tbl:underxy} 
\end{table}
\begin{table}[ht]
\renewcommand{\arraystretch}{1.5}
\caption{Transformations under $\sigma_d'$ and $\sigma_d''$ } 
\centering 
\begin{tabular}{c | c  | c c c c c c} 
\hline\hline 
 & & $|E_x|^2$ & $|E_y|^2$ & $|E_z|^2$ & $Re\{E_x^*E_y\} $ & $Re\{E_x^*E_z\} $ & $Re\{E_y^*E_z\} $ \\
 \hline
$\chi(\sigma_d')=\pm 1$ & under $\sigma_d'$ &  $|E_y|^2$ & $|E_x|^2$ & $|E_z|^2$  & $Re\{E_x^*E_y\}$ & $Re\{E_y^*E_z\}$ & $Re\{E_x^*E_z\}$ \\ 
\hline
$\chi(\sigma_d'')=\pm 1$  & under $\sigma_d''$ &  $|E_y|^2$ & $|E_x|^2$ & $|E_z|^2$  & $Re\{E_x^*E_y\}$ & $-Re\{E_y^*E_z\}$ & $-Re\{E_x^*E_z\}$ \\ 
\hline\hline
 & & $S_{xx}$ & $S_{yy}$ & $S_{zz}$ & $S_{xy}$ & $S_{xz}$ & $S_{yz}$\\
\hline
$\chi(\sigma_d')=1$& under $\sigma_d'$ & $S_{yy}$ & $S_{xx}$ & $S_{zz}$ & $S_{xy}$ & $S_{yz}$ & $S_{xz}$ \\ 
\hline
$\chi(\sigma_d')=-1$ & under $\sigma_d'$ & $-S_{yy}$ & $-S_{xx}$ & $-S_{zz}$ & $-S_{xy}$ & $-S_{yz}$ & $-S_{xz}$ \\ 
  \hline
$\chi(\sigma_d'')=1$ & under $\sigma_d''$ & $S_{yy}$ & $S_{xx}$ & $S_{zz}$ & $S_{xy}$ & $-S_{yz}$ & $-S_{xz}$ \\ 
\hline
$\chi(\sigma_d'')=-1$  & under $\sigma_d''$ & $-S_{yy}$ & $-S_{xx}$ & $-S_{zz}$ & $-S_{xy}$ & $S_{yz}$ & $S_{xz}$\\ 
\hline 
\end{tabular}
\label{tbl:underdd} 
\end{table}
\begin{table}[ht]
\renewcommand{\arraystretch}{1.5}
\caption{Transformations under $C_2$} 
\centering 
\begin{tabular}{c | c  | c c c c c c} 
\hline\hline 
 & & $|E_x|^2$ & $|E_y|^2$ & $|E_z|^2$ & $Re\{E_x^*E_y\}$ & $Re\{E_x^*E_z\}$ & $Re\{E_y^*E_z\}$ \\
 \hline
$\chi(C_2)=\pm 1$ & under $C_2$ & e & e & e & e & o & o \\ 
\hline\hline
 & & $S_{xx}$ & $S_{yy}$ & $S_{zz}$ & $S_{xy}$ & $S_{xz}$ & $S_{yz}$\\
\hline
$\chi(C_2)=1$ & under $C_2$ & e & e & e & e & o & o \\ 
\hline
$\chi(C_2)=-1$ & under $C_2$ & o & o & o & o & e & e \\ 
\hline 
\end{tabular}
\label{tbl:underc2} 
\end{table}
Let us start the analysis for the symmetry of the terms in Eq.(\ref{Eq:intterm}) in the structure with $C_{4v}$ symmetry. When both the mechanical mode and optical mode are 1-d representations, we can use \{2$\sigma_v$, $2\sigma_d$\} to analyze. According to TABLE \ref{tbl:underxy} and \ref{tbl:underdd}, for mechanical $B_1$ and $B_2$ modes and an arbitrary 1-d optical mode, each term in Eq.(\ref{Eq:intterm}) is odd under either $2\sigma_v$ or $2\sigma_d$, which leads to a zero coupling for both cubic and hexagonal crystals; for the mechanical $A_2$ mode, $f_{16}$ is even under any of \{2$\sigma_v$, $2\sigma_d$\}, while the others are odd under $2\sigma_v$, so we are left with $\int p_{16}'f_{16}\,\mathrm{d} V$. Note that in hexagonal crystal, we have $p_{16}=0$, and in cubic crystal, $p_{16}$ is periodic in $\theta$ and it is zero when $\theta$ is a multiple of $\pi/2$. Thus the coupling between a mechanical $A_2$ mode and a 1-d optical mode can be non-zero in cubic crystal when $\theta$ is not a multiple of $\pi/2$. When a mechanical BIC couples with the 2-d optical mode $E$, for the same reason as introduced in Sec. III A in the main text we can only utilize $C_2$. According to TABLE \ref{tbl:underc2}, all terms in Eq.(\ref{Eq:intterm}) are even under $C_2$ if $\chi(C_2)=1$, which is true for $A_2$, $B_1$ and $B_2$. Therefore, generally, the coupling between a mechanical BIC and a 2-d optical mode is non-zero.

Let us proceed to the triangular lattice. For the mechanical BICs with $\chi(C_2)=-1$, the volume integration of each term in Eq.(\ref{Eq:intterm}) is 0. Hence the coupling between a mechanical $B_1$ or $B_2$ mode with any optical mode is zero; while for $A_2$ and $E_2$, whose $\chi(C_2)=1$, those terms are even. Therefore, the coupling between an $E_2$ mechanical mode and any optical mode as well as the coupling between an $A_2$ mechanical mode and a 2-d optical mode are generally non-zero. For $A_2$, whose $\chi(\sigma_x)=\chi(\sigma_y)=-1$, if it couples with an1-d optical mode, we find that except $f_{16}$, all terms are odd under at least one of $\sigma_x$ and $\sigma_y$, so we are again left with $\int p_{16}'f_{16}\,\mathrm{d} V$, which leads to the same conclusion as earlier. 

\section{Simulation of an optomechanical structure with $C_{6v}$ symmetry}
\label{App:C}

\begin{figure}[!htb]
\begin{center}
\includegraphics[width=0.3\columnwidth]{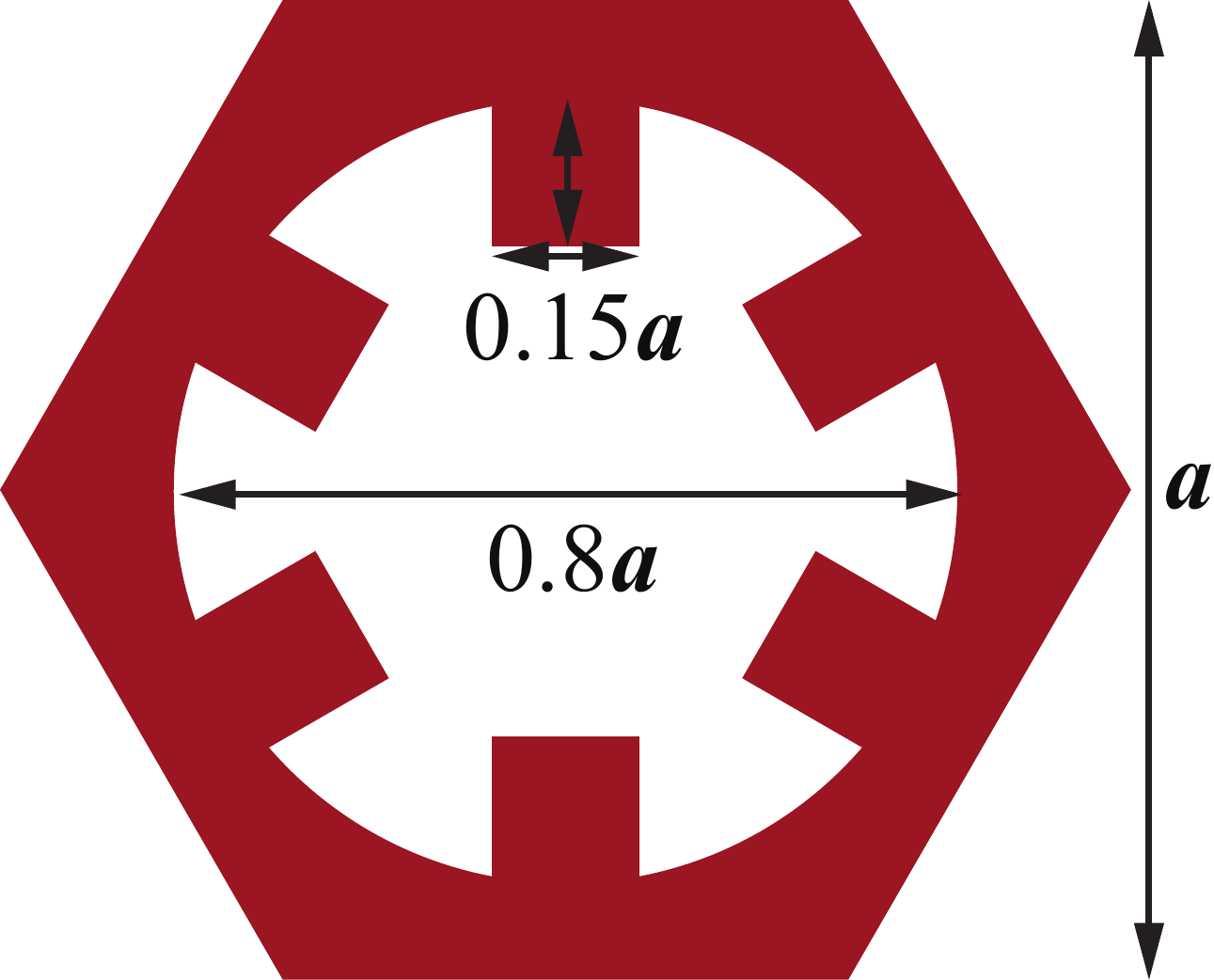}
\caption{Top view of the unit cell of an optomechanical crystal with $C_{6v}$ symmetry. In the simulation, the lattice constant $a=1\ \mu$m.}
\label{fig:triangular_unitcell}
\end{center}
\end{figure}

\begin{figure}[!htb]
\begin{center}
\includegraphics[width=0.5\columnwidth]{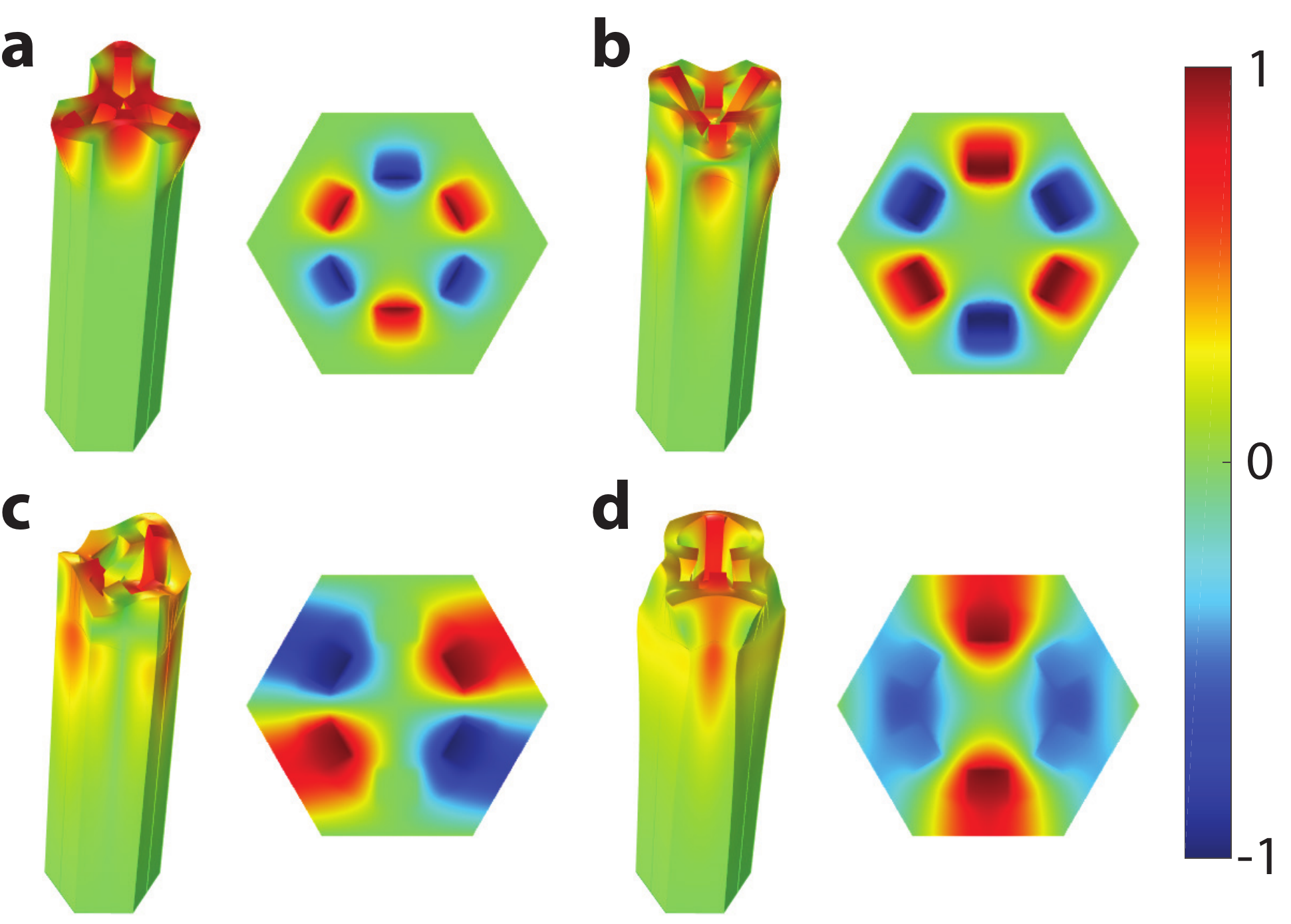}    
\caption{Mechanical modes at $\Gamma$ point. Total displacement (left) and $z-$component of the displacement at the interface between the slab and substrate (right) of the four mechanical BICs with frequency 2.69 GHz (\textbf{a}), 3.90 GHz (\textbf{b}) and 3.95 GHz (1,2) (\textbf{c},\textbf{d}).}
\label{fig:phononicbictri}
\end{center}
\end{figure}

\begin{figure}[!htb]
\begin{center}
\includegraphics[width=1\columnwidth]{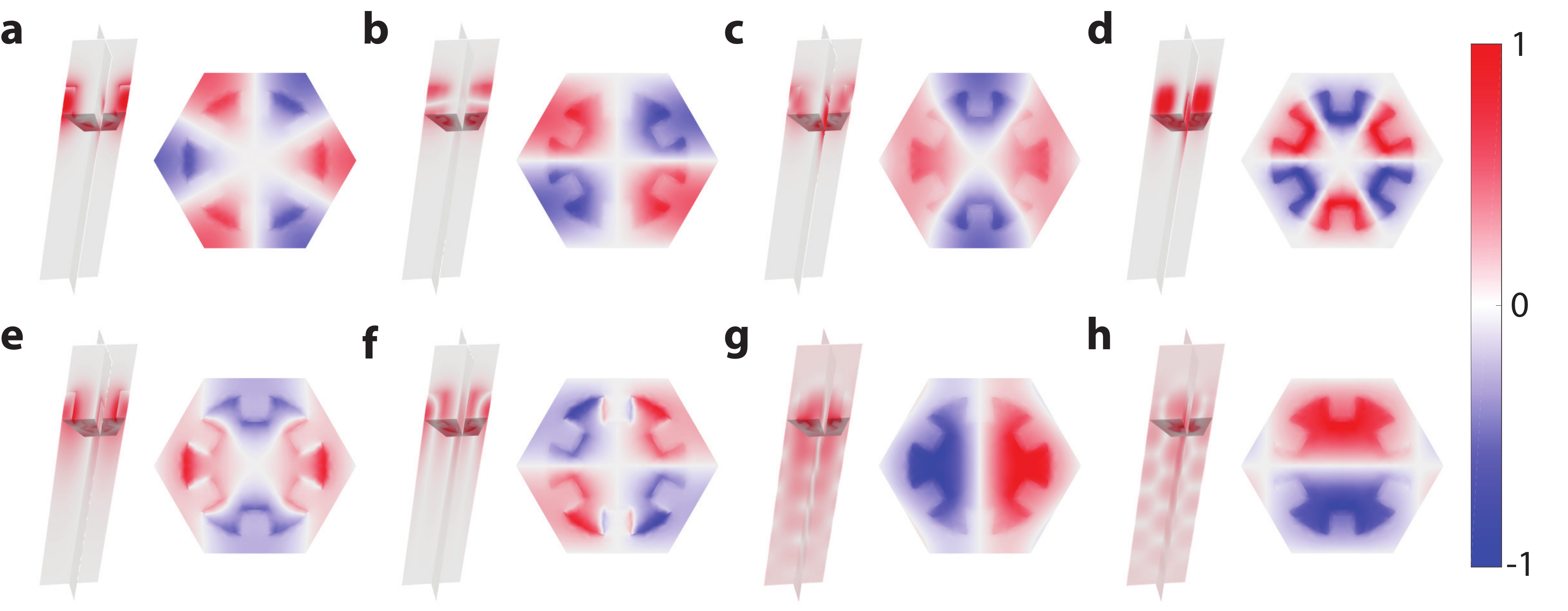}    
\caption{Optical modes at $\Gamma$ point. $|\vec{E}|^2$ (left) and $E_z$ (right) of the six optical BICs with frequency 198 THz (\textbf{a}), 202 THz (1,2) (\textbf{b},\textbf{c}), 209 THz (\textbf{d}), 222 THz (1,2) (\textbf{e},\textbf{f}) and a pair of degenerate guided resonances with frequency 226 THz (\textbf{g},\textbf{h}).}
\label{fig:photonicbictri}
\end{center}
\end{figure}

Here we show the simulation results of a triangular lattice. The material system is the same as the structure with a square lattice The top view of its unit cell is shown in Fig. \ref{fig:triangular_unitcell}. For the triangular lattice, the cut-off frequencies are $\frac{c_T}{\sqrt{3}a/2}=$3.98 GHz and $\frac{c}{n_{\textrm{SiO}_2}\sqrt{3}a/2}=$ 226 THz. Below the cut-off frequency, there are four mechanical BICs, with frequency and representation (2.69 GHz, $B_2$), (3.90 GHz, $B_2$), and (3.95 GHz (1, 2), $E_2$), respectively. There are six optical BICs, with frequency and representation (198 THz, $B_1$), (202 THz (1,2), $E_2$), (209 THz, $B_2$), and (222 THz (1,2),  $E_2$), respectively. The mode profiles of the mechanical BICs, optical BICs and optical guided resonances are shown in Fig. \ref{fig:phononicbictri} and Fig. \ref{fig:photonicbictri}, respectively. Table \ref{tbl:mbtriangle} shows the moving boundary contribution and photo-elastic contribution to the optomechanical coupling between one of the degenerate $E_2$ mechanical BIC and the optical resonances. The coupling between the other $E_2$ BIC and optical resonances happens to be zero because of the special linear combination of the degenerate modes COMSOL chooses. Optomechanical interactions between other modes are all zero as expected.

\begin{table}[H]
\caption{Optomechanical couplings in a unit cell of the $C_{6v}$ cross-structure optomechanical crystal} 
\centering 
\renewcommand{\arraystretch}{1.5}
\begin{tabular}{c|c|c|c|c|c}   
\hline\hline 
 ($\bar g_{0,\textrm{MB}}/2\pi,\ \bar g_{0,\textrm{PE}}/2\pi$) (Hz) & 198 THz ($B_1$)   & 209 THz ($B_2$) & 202 THz (1,2) ($E_2$) & 222 THz (1,2) ($E_2$) & 226 THz (1,2) ($E_1$)\\
\hline 
 3.95 GHz (2) ($E_2$) & (1.9, 0.065)  & (0.9, 0.02) & $(2.1\times 10^{4}, -1.0\times 10^{3})$ & $(3.3\times 10^3, -7.2\times 10^{2})$ & $(4.3\times 10^3, -1.1\times 10^{3})$  \\
\hline 
\end{tabular}
\label{tbl:mbtriangle} 
\end{table}

\end{document}